\documentclass[aps,pre,twocolumn,groupedaddress]{revtex4}
\usepackage{graphics} 
\usepackage{graphicx}
\newcommand{\be}{\begin{equation}}
\newcommand{\ee}{\end{equation}}
\newcommand{\bea}{\begin{eqnarray}}
\newcommand{\eea}{\end{eqnarray}}

\begin{document}


\title{Clustering Phase Transitions and Hysteresis: Pitfalls in
  Constructing Network Ensembles}


\author{David Foster} \email[]{ventres@gmail.com}
\affiliation{Complexity Science Group, University of Calgary, Calgary
  T2N 1N4, Canada }

\author{Jacob Foster} \affiliation{Complexity Science Group,
  University of Calgary, Calgary T2N 1N4, Canada }

\author{Maya Paczuski} \affiliation{Complexity Science Group,
  University of Calgary, Calgary T2N 1N4, Canada }

\author{Peter Grassberger} \affiliation{Complexity Science Group,
  University of Calgary, Calgary T2N 1N4, Canada } \affiliation{NIC,
  Forschungszentrum J\"ulich, D-52425 J\"ulich, Germany}


\date{\today}

\begin{abstract}
Ensembles of networks are used as null models in many applications.
However, simple null models often show much less clustering than their
real-world counterparts. In this paper, we study a model
where clustering is enhanced by means of a fugacity term as in the
Strauss (or ``triangle") model, but where the degree sequence is
strictly preserved -- thus maintaining the quenched heterogeneity of 
nodes found in the original degree sequence. Similar models had been 
proposed previously in
[R. Milo {\it et al.}, Science {\bf 298}, 824 (2002)].  We find that
our model exhibits phase transitions as the fugacity is changed.  For
regular graphs (identical degrees for all nodes) with degree $k>2$ we
find a single first order transition. For all non-regular networks
that we studied (including Erd\"os - R\'enyi and scale-free networks)
we find multiple jumps resembling first order transitions, together
with strong hysteresis. The latter transitions are driven by the
sudden emergence of ``cluster cores": groups of highly interconnected
nodes with higher than average degrees. To study these cluster cores
visually, we introduce {\it q-clique adjacency plots}. We find that 
these cluster cores constitute distinct communities which emerge 
spontaneously from the triangle generating process.  Finally, we
point out that cluster cores produce pitfalls when using the present
(and similar) models as null models for strongly clustered networks,
due to the very strong hysteresis which effectively leads to broken
ergodicity on realistic time scales.
\end{abstract}


\maketitle

\section{Introduction}
\label{sec:intro}

Networks are an essential tool for modeling complex systems. The nodes
of a network represent the components of the system and the links
between nodes represent interactions between those
components. Networks have been applied fruitfully to a wide variety of
social \cite{guimer_self-similar_2003, newman_social_2003},
technological \cite{broder_graph_2000}, and biological
\cite{jeong_large-scale_2000} systems. Many network properties have
been studied to discover how different functional or generative
constraints on the network influence the network's structure. In this
paper we examine five properties of particular importance: the degree
sequence \cite{newman_random_2001}, which counts the number of nodes
in the network with $k$ links; the clustering coefficient
\cite{newman_properties_2003}, which measures the tendency of
connected triples of nodes to form triangles; the number of
$q$-cliques, i.e. complete subgraphs with $q$ nodes; the assortativity
\cite{newman_assortative_2002}, which measures the tendency of nodes
to connect to other nodes of similar degree; and the modularity
\cite{newman_finding_2004}, which measures the tendency of nodes in
the network to form tightly interconnected communities. Their formal
definitions are recalled in Sec.~2.

Models of network {\it ensembles} are of interest because they
formalize and guide our expectations about real-world networks and their properties
\cite{foster_link_2007}. The most famous are the Erd\"os-R\'enyi model
of random networks \cite{erdos_random_1959}, and the scale-free
Barab\'asi-Albert model \cite{barabasi-albert_1999}. Comparison with an
{\it a priori} realistic ``null" model can also indicate which features of a
real network are expected based on the null model features, and which are surprising and thus of
interest, as in motif search \cite{milo_network_2002}. In the latter
context, the most popular ensemble is the {\it configuration model}
\cite{molloy-reed_1995} and related variants 
\cite{maslov_specificity_2002}, in which all networks with a given
number of nodes and a given degree sequence have the same weight. One
problem of the configuration model is that it shows far too little clustering; this problem is especially important when the model is applied to motif search in,
e.g., protein interaction networks \cite{baskerville_2007}.

A model where clustering can be enhanced by means of a fugacity term
in a network Hamiltonian was introduced by D. Strauss
\cite{strauss_general_1986} and studied in detail in
\cite{park_solution_2005}. In the Strauss model, the density of edges
is also controlled by a second fugacity. Thus it is a generalization
of the Erd\"os-R\'enyi model with fixed edge {\it probability}, not
with fixed edge number. In the Strauss model there is a strong first
order phase transition \cite{park_solution_2005} from a phase with
weak clustering to a phase where nearly all edges condensate in a
single densely connected cluster consisting of high degree nodes. This
phase transition is often seen as a flaw, as it does not allow the
intermediate clustering observed in most real networks \footnote{This
  transition can also be seen as first order {\it percolation}
  transition, since a giant percolating cluster is formed when $\beta$
  is increased through the critical point. It is, however, very
  different from ``explosive percolation" in Achlioptas processes
  \cite{achlioptas_2009}, which is also a first order percolation
  transition. While the Strauss model is a genuine thermodynamic model
  with Hamiltonian structure, and the phase transition happens as a
  true control parameter is increased, explosive percolation is a
  strictly nonequilibrium process where the control is done via a {\it
    density} (of established bonds). We could also try to control the
  bond density in the Strauss model, but then we would get a
  continuous transition with phase coexistence, as in any system which undergoes a
  thermodynamic first order transition.}.

In the present paper, we introduce and analyze the {\it Biased
Rewiring Model} (BRM).  As in the configuration model, we fix the
exact degree sequence--accounting for quenched heterogeneity in node properties. But as in the Strauss model, we control the
average number of closed triangles by a Hamiltonian
\cite{park_statistical_2004} containing a conjugate fugacity $\beta$.
By fixing the degree sequence we prevent the extreme condensation of
edges typical of the Strauss model, and we might {\it a priori} hope
to achieve a smooth control of the clustering. Indeed, a very similar
model, but with a slightly different Hamiltonian, had been proposed in
\cite{milo_network_2002}.

To our surprise we found that this is not the case, and the clustering
cannot be smoothly controlled. To search for phase transitions, we
plotted several characteristics (number of triangles, number of
$q$-cliques with $q=4$ and 5, assortativity, and modularity) against
$\beta$. In all these plots and for all non-regular graphs
(i.e. graphs with a non-trivial degree distribution) we found {\it
several} jumps which look like first order phase transitions (or
large Barkhausen jumps in ferromagnets 
\cite{Sears-Zemanski,Perkovic_1995,Zapperi_1997}. Associated with these 
jumps are important hysteresis effects. Further, we found that high degree
nodes play a crucial role in generating these phase transitions. It is
thus not surprising that a somewhat simpler scenario holds for regular
graphs (same degree $k$ for all nodes), where we found a single phase
transition for all $k>2$. The only case where we found no transition
at all is that of regular graphs with $k=2$. Unfortunately it is only
in the last, somewhat trivial, case that we can do exact analytic
calculations. In all other cases our results are based on simulations.

In \cite{milo_network_2002}, the Hamiltonian was chosen to bias not
towards a {\it larger} number of triangles, but towards a {\it
specific} number. In order to achieve this reliably, one needs a
fugacity which is larger than that in the BRM. In the limit of large
fugacities this is 
similar to a model with a hard constraint. In general, statistical
models with hard constraints show slower relaxation and worse ergodic
behavior than models with soft constraints \cite{barkema_MC}. We
expect thus that hysteresis effects might be even more pronounced in
the model of \cite{milo_network_2002} and might render it less useful
as a null model, even if the problem of phase transitions is
hidden. For simplicity we shall in the following call the model of
\cite{milo_network_2002} ``triangle conserving", although the name is
not strictly correct. We find that for triangle conserving rewiring,
important structures remain largely unchanged on extremely long time 
scales, requiring particular care when using the method. In general, phase transitions, strong hysteresis, and persistent structures of highly connected nodes together present substantial pitfalls for null-models of clustered networks.


In the next section we shall collect some basic background
information, including the precise definitions of the model with
unbiased rewiring and the Strauss model.  The definition of the BRM
and our numerical procedure is given in Sec.~II.F. Our main results are
found in Secs.~III.A to III.C, while some results for the model with hard
constraints of Milo {\it et al.} \cite{milo_network_2002} are
presented in Sec.~III.D Finally, Sec.~IV contains our conclusions.

\section{Background}
\label{sec:bg}

\subsection{Degree sequences}
\label{sec:deg}

The degree of a node is the number of links in which the node
participates. The network's degree sequence $\{n_k |\; k=0,1\ldots
k_{\rm max}\}$ counts the number of nodes in the network which degree
$k$. The networks studied in this paper are regular ($n_k =
\delta_{k,k_0}$), Erd\"os-R\'enyi (poissonian $n_k$), and several real
world networks with fat tails. Network properties often depend
strongly on the degree sequence \cite{newman_random_2001}.  Thus real
networks are often compared with null models which preserve the degree
sequence.

\subsection{Clustering coefficient and $q$-cliques}
\label{sec:c}

Three nodes are {\it connected}, if at least two of the three possible
links between them exits. If all three links exist, they form a {\it
  triangle}. The clustering coefficient \cite{watts_collective_1998}
measures the ``transitivity" of relationships in the network, i.e. the
probability that three connected nodes are also a triangle. Denoting
the number of triangles by $n_\Delta$ and the degree of node $i$ by
$k_i$, one has 
\be 
C = {3n_\Delta\over {1\over 2}\sum_{i=1}^N (k_i-1)k_i}.  \label{clust} 
\ee 

If every relationship in the network is transitive, $C = 1$; if no
relationships are transitive, $C = 0$. Note that the denominator of
equation \ref{clust} depends only on the degree sequence, and thus $C
\propto n_\Delta$ in any ensemble with fixed degrees.

In addition to $C$, we can also define similar higher order clustering
coefficients based on $q$-cliques, i.e. on complete subgraphs with $q$
nodes, as
\be
      C_q = {q\;n_{q-\rm clique} \over \sum_{i=1}^N {k_i \choose q-1}},
\ee
where $n_{q-\rm clique}$ is the number of $q-$cliques in the
network. Notice that $C=C_3$.  As we shall see, we can use any $C_q$
as an order parameter in the phase transitions discussed below.

\subsection{Assortativity}
\label{sec:r}

The assortativity $r$ measures the tendency for nodes in the network
to be linked to other nodes of a similar degree. It is defined as the
Pearson correlation coefficient between the degrees of nodes which are
joined by a link \cite{newman_assortative_2002}.
\be
   r = \frac{L\sum_{i=1}^L j_i k_i - [\sum_{i=1}^L j_i]^2}
              {L\sum_{i=1}^L j_i^2 - [\sum_{i=1}^L j_i]^2}
\label{assort}
\ee
Here $L$ is the number of links in the network and $j_i$ and $k_i$ are
the degrees of nodes at each end of link $i$. Thus, if high degree
nodes are linked exclusively to other high degree nodes, $r \approx
1$. If high degree nodes are exclusively linked to low degree nodes,
$r \approx -1$.


\subsection{Modularity}
\label{sec:m}

There are many methods for identifying community structure in complex
networks \cite{porter_communities}, each with its own strengths and
drawbacks. We shall use a measure proposed by Newman and Girvan
\cite{newman_finding_2004} called {\it modularity}. Assume one has a
given partition of the network into $k$ non-overlapping
communities. Define $e_{ij}$ as the fraction of all edges which
connect a node in community $i$ to a node in community $j$. Thus $a_i
= \sum_j e_{ij}$ is the fraction of all links which connect to
community $i$.  The modularity of the partition is then defined as:
\be
   Q = \sum_i (e_{ii} - a_i^2),
\label{mod}
\ee
and the modularity of the network is the maximum of $Q$ over all
partitions.  $Q$ measures the fraction of `internal' links, versus the
fraction expected for a random network with the same degree
sequence. It is large when communities are largely isolated with few
cross links.

The main problem in computing $Q$ for a network is the optimization
over all partitions, which is usually done with some heuristics. The
heuristics used in the present paper is a greedy algorithm introduced
by Newman \cite{newman_fast_2004}. We start with each node in its own
community (i.e., all communities are of size 1). Joining two
communities $i$ and $j$ would produce a change $\delta Q_{ij}$. All
pairs $(i,j)$ are checked, and the pair with the largest $\delta
Q_{ij}$ is joined. This is repeated until all $\delta Q_{ij}$ are
negative, i.e. until $Q$ is locally maximal. We follow the efficient
implementation of this method described by Clauset et al
\cite{clauset_finding_2004}.

\subsection{Exponential Network Ensembles and Network Hamiltonians}

Let us assume that ${\cal G}$ is a set of graphs (e.g. the set of all
graphs with fixed number $N$ of nodes, or with fixed $N$ and fixed
number $L$ of links, or with fixed $N$ and fixed degree sequence,
...), and $G\in {\cal G}$. Following \cite{park_statistical_2004}, a
network {\it Hamiltonian} $H(G)$ is any function defined on ${\cal
  G}$, used to define an exponential ensemble (analogous to a
canonical ensemble in statistical mechanics) by assigning a weight
\be
   P(G) \propto e^{-H(G)}
\ee
to any graph, similar to the Boltzmann-Gibbs weight.

Examples of exponential ensembles are the Erd\"os-R\'enyi model
$G(N,p)$ where $H = -L\ln [p/(1-p)]$ and the Strauss model with
\be
   H_{\rm Strauss} = \theta L - \beta n_\Delta.
\ee
Here, $p$ (which is not to be confused with $P(G)$) is the probability
that a link exists between any two nodes, while $\theta$ and $\beta$
are ``fugacities" conjugate to $L$ and $n_\Delta$, respectively.

In the configuration model, ${\cal G}$ is the set of all graphs with a
fixed degree sequence and $H=0$. Thus all graphs have the same
weight. In contrast, in the ``triangle conserving" biased model of 
Milo {\it et al.} \cite{milo_network_2002} ${\cal G}$ is again the 
set of graphs with fixed degree sequence, but
\be
   H_{\rm Milo} = \beta |n_\Delta - n_{\Delta,0}|. 
\ee
where $n_{\Delta,0}$ is some target number of triangles, usually the
number found in an empirical network.  Finally, in the BRM, ${\cal G}$
is again the same but
\be
   H_{\rm BRM} = - \beta n_\Delta.
\ee
Thus, while large weights are given in the BRM (with $\beta >0$) to
graphs with many triangles (high clustering), in the model of
\cite{milo_network_2002} the largest weights are given to graphs with
$n_\Delta = n_{\Delta,0}$.

\subsection{Simulations: Rewiring}
\label{sec:rewire}

Simulations of these ensembles are most easily done by the Markov
chain Metropolis-Hastings method \cite{barkema_MC}. This is
particularly easy for models without fixed degree sequences,
e.g. the Strauss model. There, new configurations are simply
generated by randomly adding or removing links. This is not possible
for the ensembles with fixed degree sequences, where the most natural
method is {\it rewiring} \cite{maslov_specificity_2002}.  We will
first discuss the unbiased case (the configuration model), and
then discuss the two biased cases $H_{\rm Milo}$ and $H_{\rm BRM}$.

\subsubsection{Unbiased Rewiring}

Starting from a current graph configuration $G$, a new graph $G'$ is
proposed as follows: Two links which have no node in common are chosen
at random, e.g. $X$---$Y$ and $W$---$Z$.  Links are then swapped
randomly either to $X$---$W$ and $Y$---$Z$, or to $X$---$Z$ and
$Y$---$W$. If this leads to a double link (i.e. one or both of the
proposed new links is already present), the new graph $G'$ is
discarded and $G$ is kept. Otherwise, $G'$ is accepted.  It is easily
seen that this conserves the degree sequence, satisfies detailed
balance, and is ergodic \cite{maslov_specificity_2002}. Thus it leads
to equidistribution among all graphs with the degree sequence of the
initial graph.

Although there seem to exist no exact results on the speed of
equilibration, previous experience 
\cite{maslov_specificity_2002,baskerville_2007} suggests that
the above unbiased rewiring is very fast indeed, and can be used
efficiently even for large networks.

\subsubsection{Biased Rewiring}

For biased rewiring with a Hamiltonian $H(G)$, the proposal stage is
the same, and only the acceptance step has to be modified, according
to the standard Metropolis-Hastings procedure
\cite{hastings_monte_1970, barkema_MC}: If $H(G') \leq H(G)$, then
$G'$ is accepted (unless it has a double link, of course). Otherwise
the swap is accepted only with a probability
\be
   p = e^{H(G) - H(G')}                                     \label{prob}
\ee
which is less than 1.

The detailed protocols for simulating the two biased models studied in
this paper are different. For the BRM we start with the actual network
$G_0$ whose degree sequence we want to use, and propose first $M_0$
{\it unbiased} swaps, with $M_0$ sufficiently large so that we end up
in the typical region of the unbiased ensemble. After that we increase
$\beta$ in small steps (typically $\delta\beta = 0.002$), starting
with $\beta=0$. After each step in $\beta$ we propose $M_1$ swaps to
equilibrate approximately, and then take take at fixed $\beta$ an
ensemble average (with further equilibration) by making $m$
measurements, each separated by $M_2$ additional proposed swaps. Thus
the total number of proposed swaps at each fixed $\beta$ is $M_1 +
(m-1)M_2$. Typically, $M_0 \approx 10^6, M_1 > 10^5, M_2 \approx 10^3
- 10^5$, and $m\approx 500 - 10,000$.

Following the $m$ measurements we increase $\beta$ and repeat
  this procedure, until a preset maximal value $\beta_{\rm max}$ is
reached. After that, we reverse the sign of $\delta\beta$ and continue
with the same parameters $M_1, M_2,$ and $m$ until we reach again
$\beta=0$, thereby forming a hysteresis loop. Fugacity values during
the ascending part of the loop will in the following be denoted by
$\beta^+$, those in the descending part as $\beta^-$. In cases where
we start from a real world network with $n_{\Delta,0}$ triangles, we
choose $\beta_{\rm max}$ sufficiently large so that
$n_{\Delta}(\beta^+) > n_{\Delta,0}$, i.e. the clustering covered by
the hysteresis loop includes the clustering coefficient of the
original network.
 
For the biased model of Milo {\it et al.} \cite{milo_network_2002} we
skip the first stage (i.e., we set $M_0=0$), and we jump immediately
to a value of $\beta$ (estimated through preliminary runs) which must
be larger than the smallest $\beta^+$ which gave rise to
$n_{\Delta,0}$ triangles in the ascending part of the loop discussed
above. We first make $M_1$ swaps to equilibrate, and then make $m$
measurements, each separated by $M_2$ further swaps (an alternative
protocol using multiple annealing periods will be discussed in
Sec.~\ref{sec:null}). Averages are taken only over configurations with
exactly $n_{\Delta,0}$ triangles. If $\beta$ is too small, the bias
will not be sufficient to keep $n_\Delta$ near $n_{\Delta,0}$, and
$n_\Delta$ will drift to smaller values. Even if this is not the case
and if $\beta$ is sufficiently large in principle, the algorithm will
slow down if $\beta$ is near its lower limit, since then $n_\Delta$
will seldom hit its target value. On the other hand, if $\beta$ is too
large then the algorithm resembles an algorithm with rigid constraint,
which usually leads to increased relaxation times. Thus choosing an
optimal $\beta$ is somewhat delicate in this model.

\section{Results}
\label{sec:results}
We explored the behavior of the BRM for three different classes of
degree sequences: Fixed $k$ networks, in which every node of the
network is degree $k$; Poisson degree distributions as in
Erd\"os-R\'enyi networks; and typical fat-tailed distributions as in
most empirical networks. Although we studied many more cases
(Erd\"os-R\'enyi networks with different connectivities and sizes and
several different protein-protein interaction networks), we present
here only results for fixed $k$ with different $k$, for one
Erd\"os-R\'enyi network, and for two empirical networks with
fat-tailed degree distributions: A high energy physics collaboration
network \cite{newman_structure_2001} and a protein-protein interaction
network for yeast ({\it S. cerevisiae})
\cite{gavin_functional_2002}). In all but fixed $k$ networks we found
multiple discontinuous phase transitions, while we found a single
phase transition in all fixed $k$ networks with $k>2$.

\subsection{Fixed $k$ networks, analytic and simulation results}
\label{sec:fixedk}

\begin{figure}
\includegraphics[scale=.32]{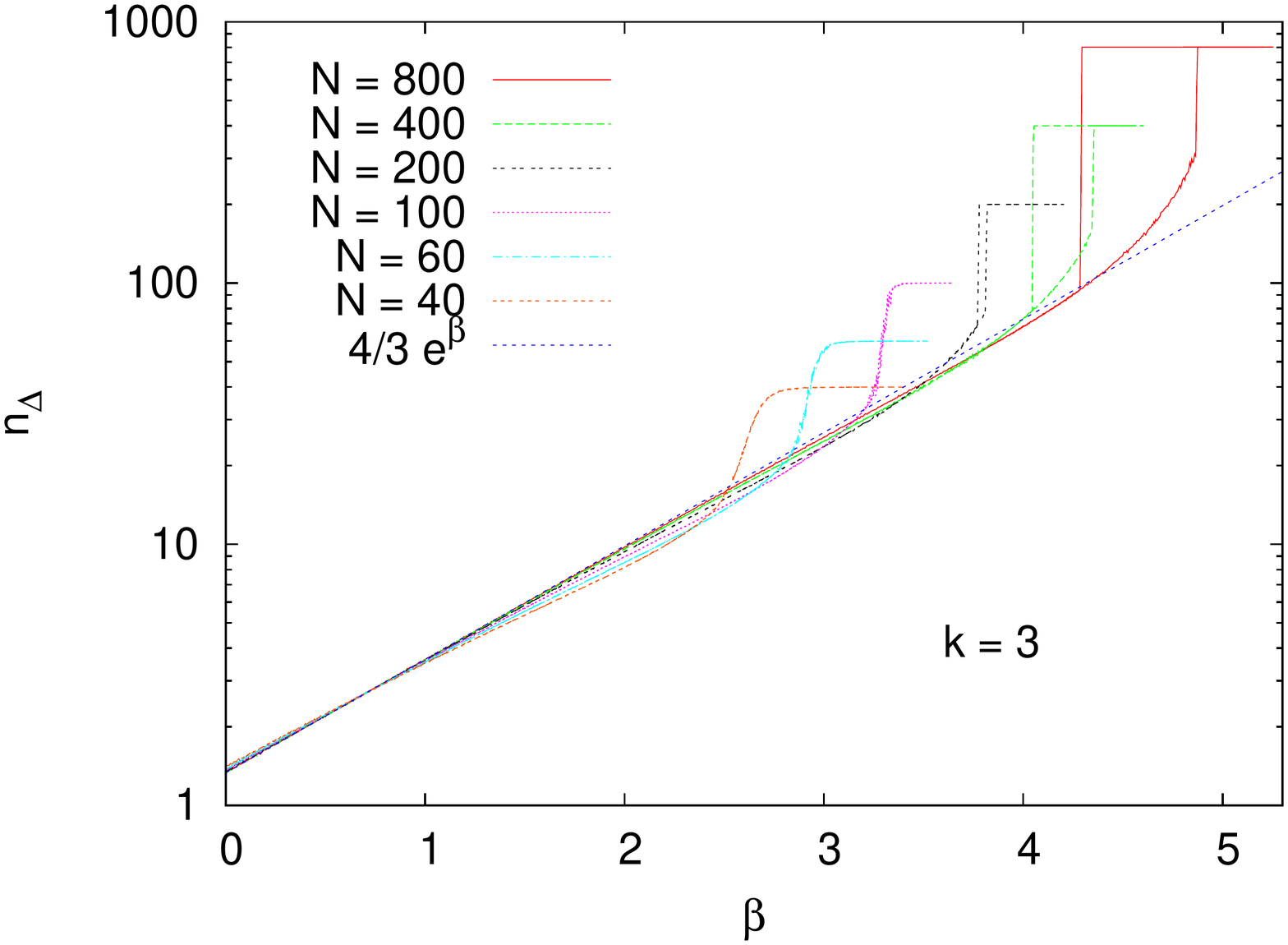}
\caption{\label{Figure1} (color online)
  Average number of triangles for networks with fixed $k=3$, plotted
  against $\beta$. All curves are obtained by full hysteresis cycles,
  with $M_1 = 200000$ initial swaps after increase/decrease of
  $\beta$, and $M_2=5000$ additional swaps after each of $m = 40000$
  measurements at the same value of $\beta$.  Hysteresis loops are
  seen for $N\geq 200$, but not for $\leq 100$. The straight line
  corresponds to the approximation Eq.~(\ref{n_approx}).}
\end{figure}

\subsubsection{Fixed $k$ simulations}

For each $k$, the configuration with maximal $n_\Delta$ is a disjoint
set of $(k+1)-$cliques, i.e. the graph decomposes into disjoint
completely connected components of $k+1$ nodes. When $N$ is divisible
by $k+1$, this gives 
\be 
n_\Delta^{(k,\rm max)} = {N\over k+1}{k+1\choose 3}.  \label{n_max} 
\ee
 For $k=2$, this $n_\Delta^{(k,\rm max)}$ is reached in a smooth
 way. For each $k\geq 3$, in contrast, and for sufficiently large $N$,
 $n_\Delta$ first increases proportional to $\exp(\beta)$, but then the
 increase accelerates and finally it jumps in a discrete step to a value
 very close to $n_\Delta^{(k,\rm max)}$. This is illustrated for $k=3$
 in Fig.~\ref{Figure1}, where we plot hysteresis curves for $n_\Delta$
 against $\beta$. From this and from similar plots for different $k$
 we observe the following features:
\begin{itemize}
\item For small $\beta$, all curves are roughly described by 
\be
   n_\Delta \approx {(k-1)^3 \over 6} e^\beta                \label{n_approx}
\ee
(see the straight line in Fig.~\ref{Figure1}), and this approximation
seems to become exact as $N\to\infty$. Notice that this implies that
$n_\Delta$ is independent of $N$, and the clustering coefficient is
proportional to $1/N$.

\item While the curves are smooth and do not show hysteresis for small
  $N$, they show both jumps and hysteresis above a $k-$dependent value
  of $N$. This is our best indication that the phenomenon is basically
  a first order phase transition, similar to the one in the Strauss
  model. Above the jump, the curves saturate (within the resolution of
  the plot) the bound given in Eq.(\ref{n_max}).

\item The critical values of $\beta$ increase logarithmically with
  $N$, although a precise determination is difficult due to the
  hysteresis. Notice that size dependent critical points are not very
  common, but there are some well known examples. Maybe the most
  important ones are models with long range or mean field type
  interactions, where the number of interaction terms increases faster
  than $N$.  In the present case the reason for the logarithmic
  increase of $\beta_c$ is that networks with fixed $k$ become more
  and more sparse as $N$ increases. Thus also the {\it density} of
  triangles (the clustering coefficient) decreases, and in a Markov
  chain MC method, there are increasingly more proposed moves which
  destroy triangles than moves which create them. To compensate for
  this and make the number of accepted moves equal, $\exp(\beta_c)$
  has to increase $\propto N$.
\end{itemize}

\begin{figure}
\includegraphics[scale=.22]{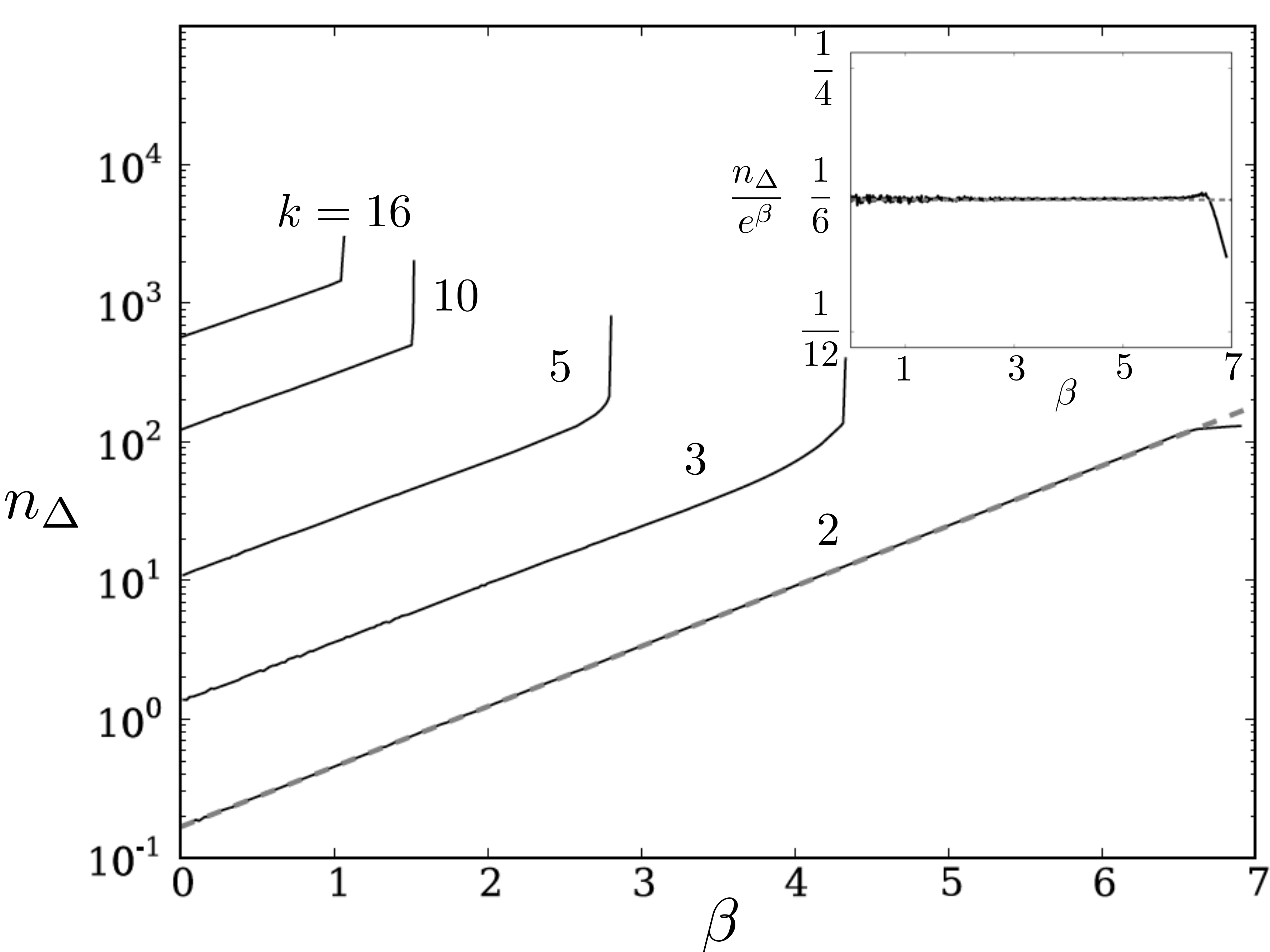}
\caption{\label{Figure2} Average number of triangles of fixed-$k$
  degree sequence networks, with $k = 2,3,5,10,$ and $16$, versus the
  fugacity (bias) $\beta$. Network size is $N=400$ for all curves. In
  these simulations $\beta$ was slowly increased, until a jump in
  $n_\Delta$ was seen (for $k\geq 3$). The straight line shows the
  theoretical prediction for $k = 2$: $n_\Delta = \frac{1}{6}
  e^{\beta}$. The inset shows $n_\Delta/e^{\beta}$ for $k=2$.}
\end{figure}

In Fig. \ref{Figure2} we show the average number of triangles as a
function of $\beta$ for fixed $k$ networks, $k = 2, 3, 5, 10,$ and
$16$, with $N = 400$ nodes. For each curve we used $M_1=4000000$
initial swaps after each increase in $\beta$, and $M_2=200000$ additional
swaps after each of $m\geq 5000$ measurements at the same value of
$\beta$.  For clarity we show only values for increasing $\beta$,
although there is strong hysteresis for all $k\geq 3$ and for $N=400$.

For $k=2$ there is not only no hysteresis, but there is indeed no
indication of any phase transition. As seen from the inset, the data
for $k=2$ are for all values of $\beta$ very well described by
Eq.~(\ref{n_approx}), up to the point where it reaches the bound
Eq.(\ref{n_max}). Close to that point there is a tiny bump in the
curve shown in the inset, that will be explained in the next
sub-sub-section.

\subsubsection{$k = 2$ analytic results}

We now give an analytical derivation of Eq.~(\ref{n_approx}) for
$k=2$, and we also show that this should become exact in the
limit $N\to\infty$.

In a fixed $k = 2$ network, there are $N$ nodes and $N$ links all
arranged in a set of disjoint simple loops. Triangles are the smallest
possible loops, since self-links and double links are not allowed. For
large $N$ and small $\beta$ nearly all loops are large, thus the
number of loops of length $<7$ is of order $1/N$ and can be neglected
for $N\to\infty$ and finite $\beta$, except that we have to allow for
a small fraction of loops to have length 3, in order to achieve
equilibration of the rewiring procedure.

Consider now a network of size $N$ with $n_\Delta$ triangles and a
triangle bias $\beta$.  The rewiring process will reach an
equilibrium, when the probability of destroying a triangle is equal to
the probability of creating a new one.

First we calculate the probabilities of randomly generating a swap
which destroys a triangle.  The total number of ways to choose a pair
of links and perform a swap is $\cal{N}$ $= \frac{N(N - 1)}{2} \times
2$, where $\frac{N(N - 1)}{2}$ gives the number of distinct pairs of
links and the extra factor of 2 accounts for the two possible ways of
swapping the links. To destroy a triangle, one of the links must be
chosen from it, and the other from a larger loop (the chance that both
links are chosen from triangles, which would lead to the destruction
of both, can be neglected). There are $3n_\Delta$ possible links in
triangles to choose from, and $(N- 3n_\Delta)$ links in larger
loops. Thus the probability of choosing a swap which would destroy a
triangle is

\be 
p_{\Delta-} = \frac{3n_\Delta(N-3n_\Delta)\times 2}{\cal{N}} =
\frac{6n_\Delta}{N} \times [1+O(N^{-1})], 
\ee

where the factor of $2$ in the numerator corresponds to the fact that
both possible swaps destroy a triangle and the correction term takes
also into account the neglected loops of lengths 4,5, and 6.

To add a triangle to the network, two links must be chosen from the
same long loop. They must be separated by exactly two links. There are
$\ell$ such pairs in a loop of length $\ell$, and thus the total
number of such pairs in the network is $N$, neglecting terms of $O(1)$,
corresponding to the triangles and loops shorter than $7$. This leaves
us with the probability of adding a triangle
\be
   p_{\Delta+} = \frac{N}{\cal{N}} = N^{-1} \times [1+O(N^{-1})].
\ee
where there is no factor of $2$ in the numerator because only one of
the two possible swaps will lead to triangle creation. Balance will be
achieved when
\be
   p_{\Delta+} = e^{-\beta}p_{\Delta-},
\ee 
giving
\be
    n_\Delta = \frac{e^\beta}{6}                         \label{balance2}
\ee 
up to correction terms of order $1/N$, which is just
Eq.~(\ref{n_approx}) for $k=2$.

The simple exponential behavior of $n_\Delta$ with $\beta$ occurs
because swaps create/destroy triangles (except in the rare case of
breaking up a loop of length 6) independently and one at a time. For
networks with nodes of degree greater than $2$ this is still basically
true when $\beta$ is small. But as $\beta$ increases, nodes cluster
together more densely, allowing each link to participate in many
triangles. For large values of $\beta$ these links, once formed,
become difficult to remove from the network. This cooperativity -- in which the presence of triangles helps other triangles to form and makes it harder for them to be removed -- explains intuitively the existence of first order phase transitions for $k\geq3$ but not for $k=2$, where the cooperative effect is not possible.

Indeed, for $n_\Delta$ very close to $n_\Delta^{(2,\rm max)}$ there is
{\it some} cooperativity even for $k=2$. The configuration with
$n_\Delta=n_\Delta^{(2,\rm max)}$ can be changed only by breaking up
{\it two} triangles and joining their links in a loop of length
6. When $n_\Delta$ is close to $n_\Delta^{(2,\rm max)}$, link swaps
which involve two triangles become increasingly prevalent. The
tendency to form and destroy triangles two at a time introduces a very
weak cooperativity, which is only strong enough to be effective when
$n_\Delta^{(2,\rm max)}-n_\Delta=O(1)$. It is thus not enough to give rise to a phase
transition, but it explains the small bump seen in the inset of
Fig.~\ref{Figure2}.

\subsection{Networks with non-trivial degree sequences}
\label{sec:ER}

\begin{table}
\begin{tabular}{ c  c  c  c  c  c l }
\multicolumn{6}{c}{Network properties} \\
\hline \hline
Network & $N$  & $\langle k\rangle $ & $C$ & $r$ & $Q$ & Comment and Ref.\\ 
\hline
ER & $800$ & $5.0$ & $.002$ & $-.0004$ & $0.196$ & Erd\"os-R\'enyi \\
HEP & $7610$ & $4.1$ & $.33$ & $.29$ & $0.397$ & scientific
collab. \cite{newman_structure_2001}\\
Yeast & $1373$ & $10.0$ & $.58$ & $.58$ & $0.380$ & protein binding
\cite{gavin_functional_2002}\\
\end{tabular}

\caption{\label{table1} The number of nodes $N$, the number of links
  $L$, the average degree $\langle k\rangle$, the clustering
  coefficient $C$, and the assortativity $r$ for each of the networks
  discussed in Sec.~\ref{sec:ER}.}
\end{table}

\begin{figure}
\includegraphics[scale=.22]{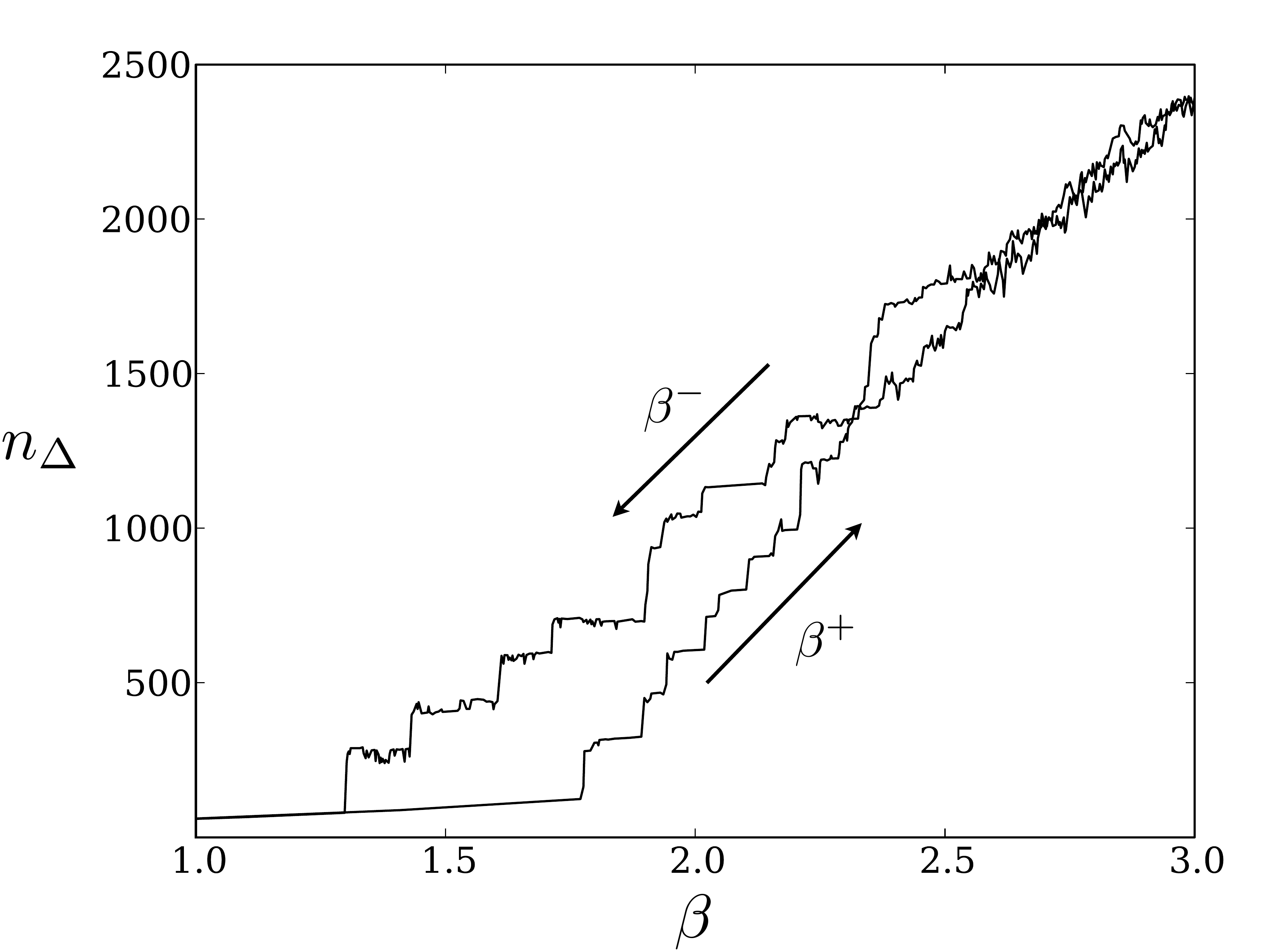}
\caption{\label{Figure3} Average number of triangles in BRM
  networks with an ER degree sequence with 800 nodes and $\langle
  k\rangle = 5$, plotted against the bias $\beta$. The lower curve
  corresponds to slowly increasing $\beta$, the upper to decreasing
  $\beta$.}
\end{figure}

\begin{figure}
\includegraphics[scale=.22]{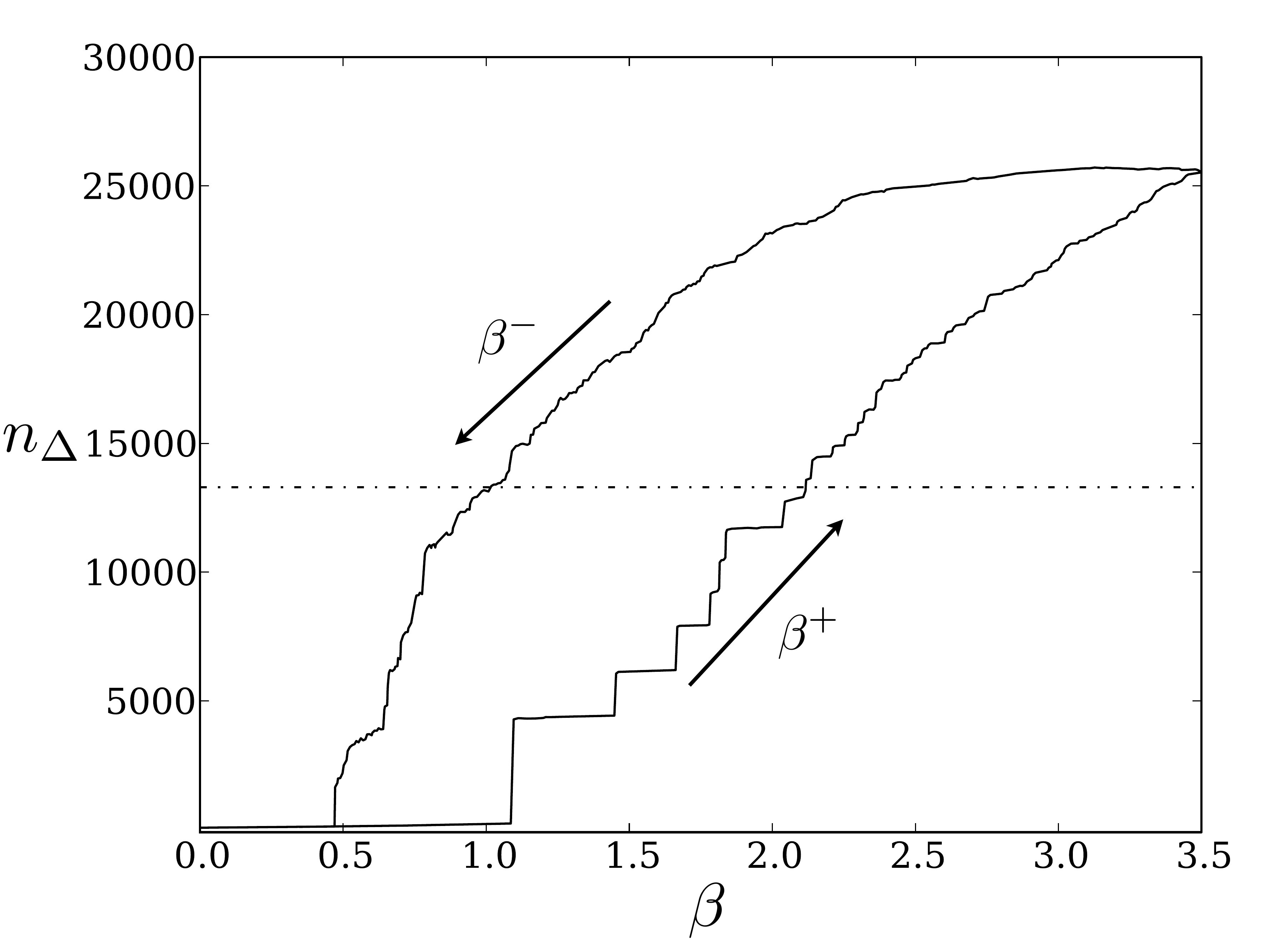}
\caption{\label{Figure4} Similar to Fig.~\ref{Figure3}, but for the
  HEP network (see Table~\ref{table1}). The dotted line indicates the number 
  of triangles in the real network.}
\end{figure}

\begin{figure}
\includegraphics[scale=.22]{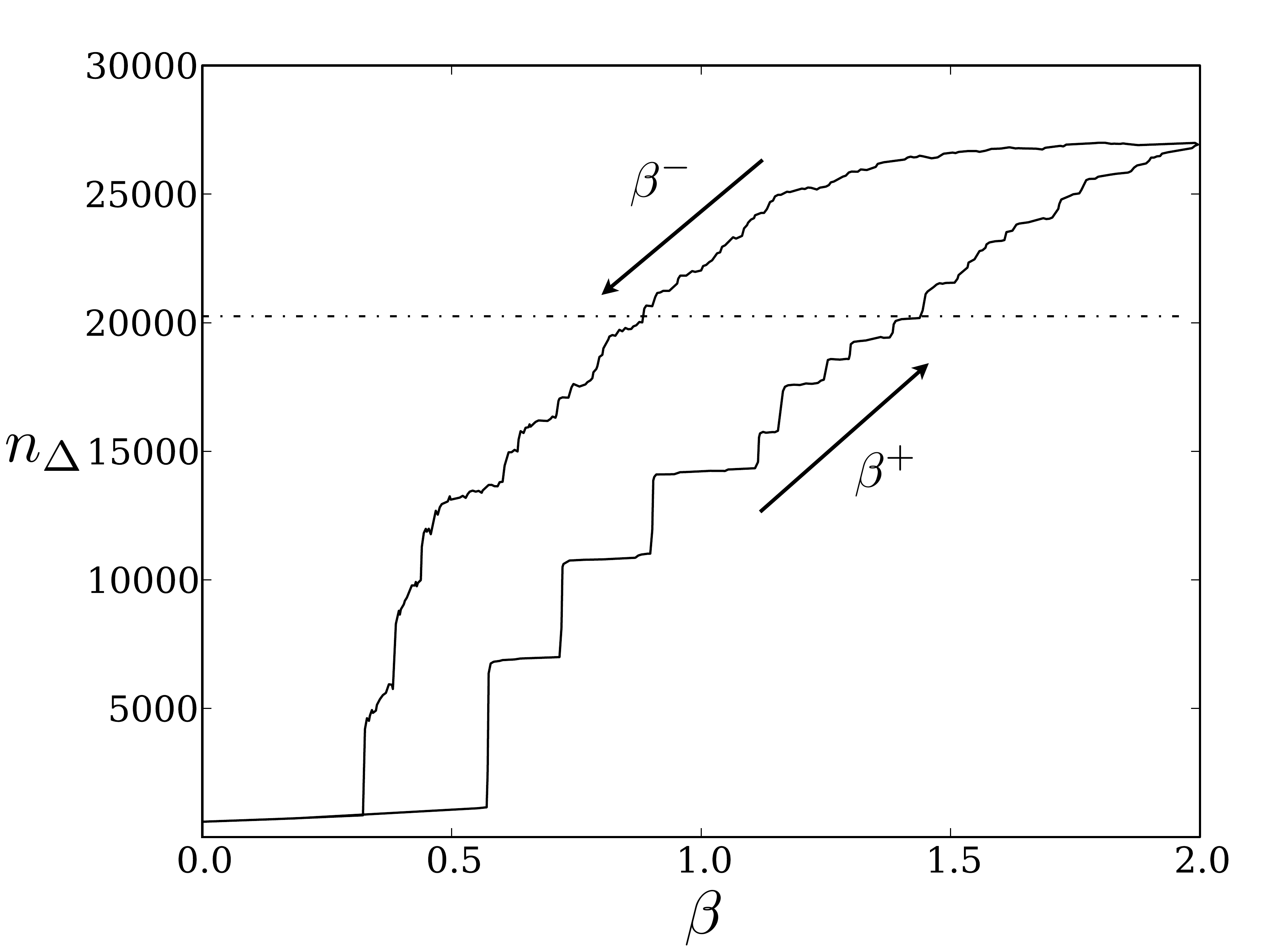}
\caption{\label{Figure5} Similar to Fig.~\ref{Figure3}, but for the
  Yeast network (see Table~\ref{table1}).}
\end{figure}

We explored the behavior of our biased rewiring model for various
degree sequences. These included Erd\"os-R\'enyi graphs
with different sizes and different connectivities and several
real-world networks. The latter typically show more or less fat
tails. In order to find any dependence on the fatness, we also changed
some of the sequences manually in order to reduce or enhance the
tails. We found no significant systematic effects beyond those visible
already from the following three typical networks, and restrict our
discussion in the following to these: an Erd\"os-R\'enyi graph
\cite{erdos_random_1959} (henceforth ER), a high energy physics
collaboration network (HEP)
\cite{newman_structure_2001}, and a yeast protein binding network
(\cite{gavin_functional_2002} (Yeast). Some of their properties
are collected in Table~\ref{table1}.

\begin{figure}
\includegraphics[scale=.22]{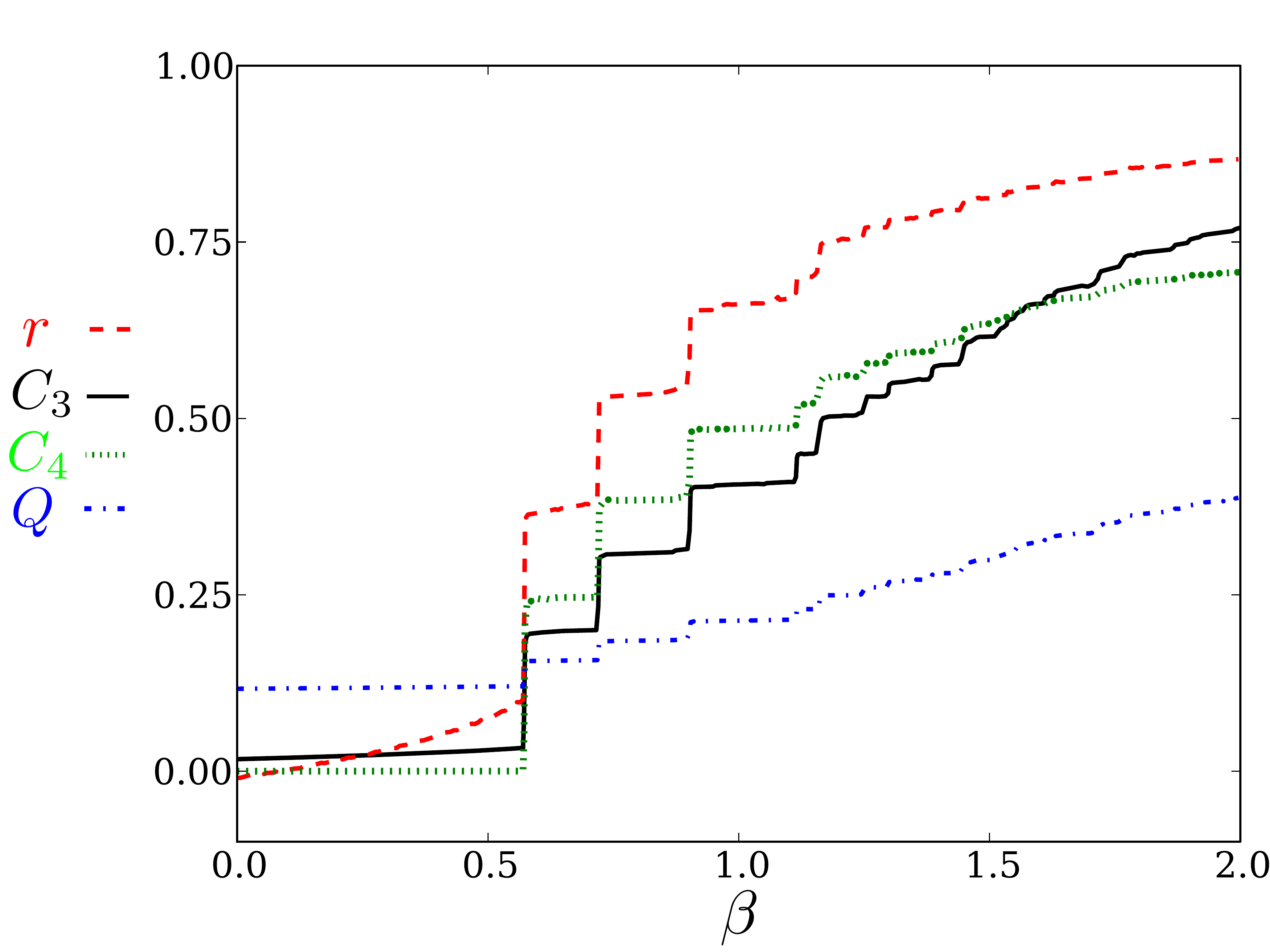}
\caption{\label{Figure6} (color online) Four network characteristics
  (modularity ($Q$), clustering coefficient ($C_3$), 4-clique
  clustering coefficient ($C_4$), and assortativity ($r$)) for BRM
  networks with the Yeast degree sequence of Table~\ref{table1} versus
  $\beta$.  These data are drawn from the same simulation as in
  Fig.~\ref{Figure5}, but for clarity only the results for
  increasing values of $\beta$ are shown.}
\end{figure}

Figs.~\ref{Figure3}, \ref{Figure4}, and \ref{Figure5} show $n_\Delta$
for these three networks.  In each case $M_0 = 10^6, M_1 = 1.5\times
10^5, M_2 = 50000$, and $m = 500$. In each of them a full hysteresis
cycle is shown, with the lower curves (labeled $\beta^+$)
corresponding to increasing and the upper curves ($\beta^-$)
corresponding to decreasing $\beta$. In Figs.~\ref{Figure4} and
\ref{Figure5} the dotted line shows the number of triangles in the
empirical networks.

For small values of $\beta^+$ all three figures exhibit a similar
exponential increase in the number of triangles as that observed in
fixed $k$ networks. At different values of $\beta$, however, there is
a sudden, dramatic increase in $n_\Delta$, which does {\it not},
however, lead to saturation as it did for fixed $k$. This first phase
transition is followed by a series of further transitions through
which the network becomes more and more clustered. Many of them are
comparable in absolute magnitude to the first jump. Although the rough
positions of the jumps depend only on the degree sequences, their
precise positions and heights change slightly with the random
number sequences used and with the speed with which $\beta$ is
increased. Thus the precise sequence of jumps has presumably no deeper
significance, but their existence and general appearance seems to be a
universal feature found in {\it all} cases.

Associated with the jumps in $n_\Delta$ are jumps in all other network
characteristics we looked at, see Fig.~\ref{Figure6}. Although the
locations of the jumps in $n_\Delta$ depend slightly on the details of
the simulation, the jumps in the other characteristics occur always at
{\it exactly} the same positions as those in $n_\Delta$. Obviously, at
each jump a significant re-structuring of the network occurs,
which affects all measurable quantities. Speculations how these
reorganizations can be best described and what is their most
``natural" driving mechanism will be given in the next subsection.

\begin{figure}
\includegraphics[scale=.32]{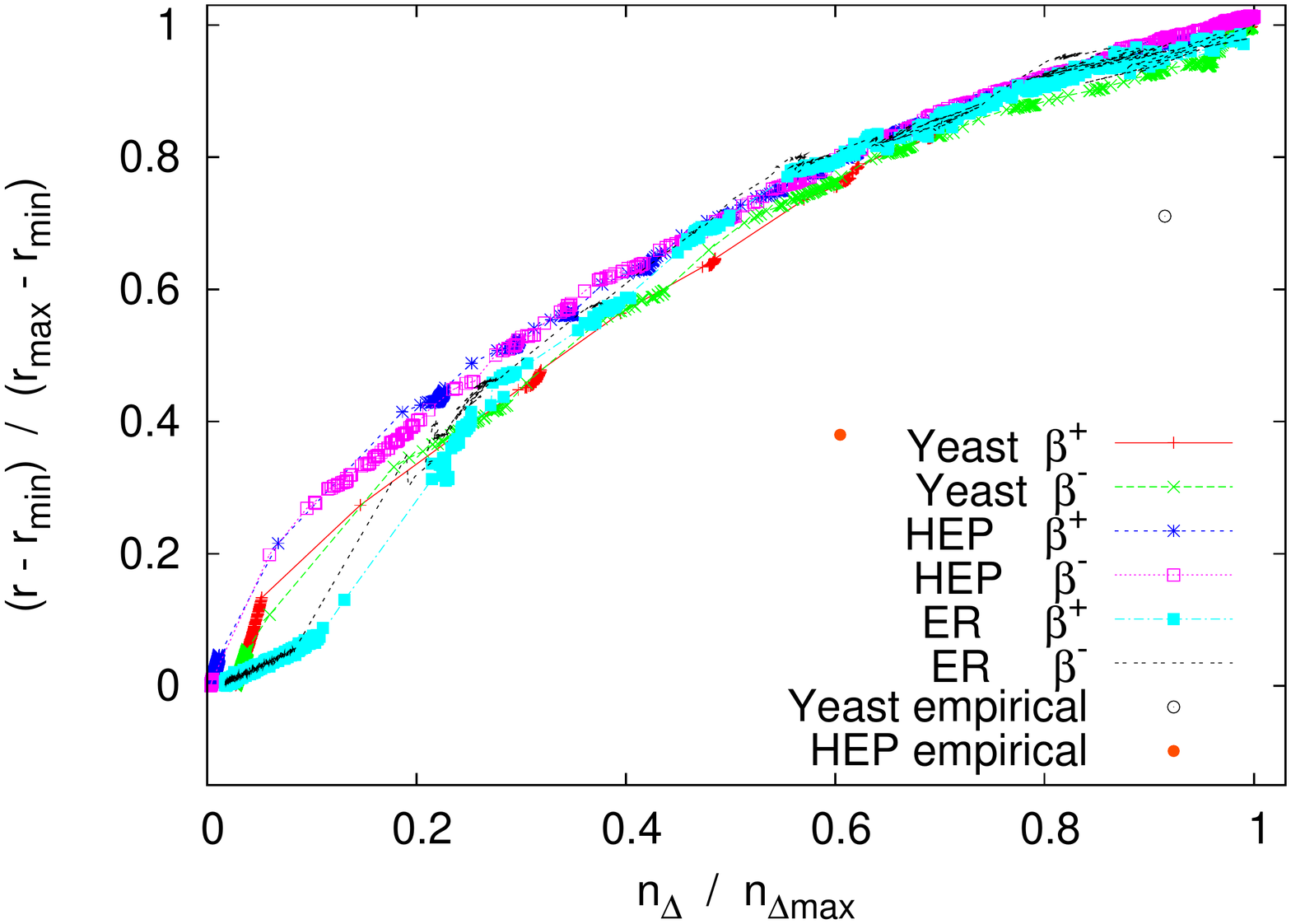}
\caption{\label{Figure7} (color online) Values of the rescaled
  characteristics $n_\Delta/n_{\Delta,\rm max}$ and $(r-r_{\rm
    min})/(r_{\rm max}-r_{\rm min})$, measured at the same values of
  $\beta^\pm$, and plotted against each other. The points represent
  the values for the real HEP and Yeast networks.}
\end{figure}

In the downward branch of the hysteresis loop, as $\beta^-$ decreases
toward zero, the number of triangles remains high for a long time,
forming a significant hysteresis loop. This loop suggests that all
jumps should be seen as discontinuous (first order) phase
transitions. Since all studied systems are finite, these hysteresis
loops would of course disappear for infinitely slow increase/decrease
of the bias. But the sampling shown involved $>25$ million attempted
swaps at each value of $\beta$, and no systematic change in the
hysteresis was seen when compared to twice as fast sweeps.

In Fig.~\ref{Figure7}, we plotted $n_\Delta$ against the assortativity
for the {\it same} values of $\beta^\pm$, normalizing both quantities
to the unit interval. The hope was that in this way we would get
universal curves which are the same for $\beta^+$ and $\beta^-$, and
maybe even across different networks. Indeed we see a quite remarkable
data collapse. It is certainly not perfect, but definitely better than
pure chance. It suggests that biasing with the BRM leads to networks
where the two characteristics $n_\Delta/n_{\Delta,\rm max}$ and
$(r-r_{\rm min})/(r_{\rm max}-r_{\rm min})$ are strongly -- but
non-linearly -- correlated. This indicates a potential scaling relationship between these network parameters in our model. For the two empirical networks, we show
also the real values of these characteristics. They fall far from the
common curve, indicating that these networks are not typical for the
BRM with any value of $\beta$.

\begin{figure*}[htp]
\includegraphics[scale=.5]{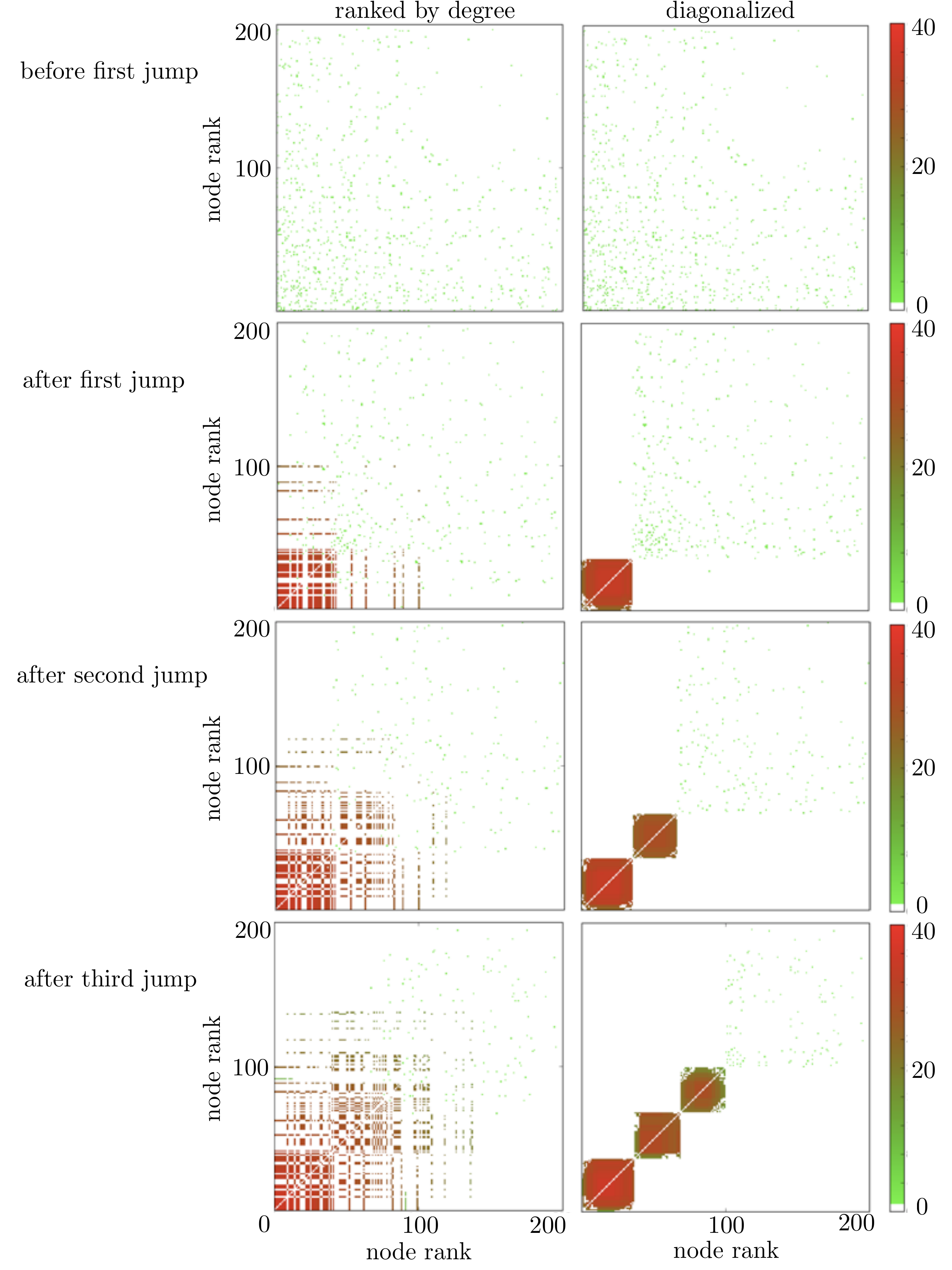}
\caption{\label{Figure8} (color online) Relevant parts of 3-clique
  adjacency plots for the Yeast degree sequence. The color of each point indicates the number of $3$cliques (or triangles) in which the link participates, as given by the scale on the right hand side. Each pair of plots shows (from top to bottom) the 3CAP for a typical member of the
  ensemble shortly before the first jump seen in Fig.~\ref{Figure5},
  shortly after it, shortly after the second jump, and shortly after
  the third jump. The plots on the left hand side show the 3CAP with
  the nodes ranked in order of their degree. In the ``diagonalized"
  plots we rearranged the ranking so that nodes which participate in
  the three clusters formed by each jump are ranked together, at the
  head of the list. The rest of the nodes are ranked by degree.}
\end{figure*}


Among the three networks studied here, the ER network is closest to a
fixed $k$ network, and it should thus show behavior closest to that
studied in the last subsection. This is not very evident from
Figs.~\ref{Figure3} to \ref{Figure5}. On the other hand, we see
clearly from these figures that the position of the first transition
-- in particular in $\beta^+$ -- decreases with the average
degree. Also, hysteresis seems to be more closely tied to individual
jumps for ER, while it is more global (and thus also more important
overall) for HEP and Yeast.

For the HEP and Yeast networks, we can compare the clustering of the
BRM ensemble to that in the real empirical networks. The latter
numbers are shown as a dashed lines in Figs.~\ref{Figure4} and
\ref{Figure5}. In both cases, the line intersects the hysteresis loop
where it is very broad.  This means that a large value of $\beta^+$ is
required to reach the real network's level of clustering when the bias
is increased, whereas a much lower value $\beta^-$ must be reached
before these triangles can be rewired out of the network again. This
gap between $\beta^+$ and $\beta^-$ at fixed $n_\Delta$ has important
implications for the triangle ``conserving" null model of
Ref.~\cite{milo_network_2002}, as we will discuss later.

\subsection{Clique adjacency plots and clustering cores}

Up to now we have not given any intuitive arguments why clustering
seems to increase in several jumps, and not in one single jump or in a
continuous way. {\it A priori} one might suggest that each jump is related
to the break-up of a connected component into disconnected subgraphs,
just as the phase transition in regular graphs was associated to such
a break-up. By counting the numbers of disconnected components we
found that this is not the case, except in special cases \footnote{If
  the degree sequence has, e.g., 20 hubs each of degree 19 and
  otherwise only nodes with degrees $<4$, then we would expect that
  the first jump leads to a clique with all 20 hubs which would then
  be disconnected from the rest. But this is a very atypical
  situation}.

Instead, we will now argue that each jump is associated with the
sudden formation of a highly connected cluster of high degree
nodes. The first jump in a scan with increasing $\beta$ occurs when
some of the strongest hubs link among themselves, forming a highly
connected cluster.  Subsequent jumps indicate the formations of other
clusters with high intra- but low inter-connections. What
distinguishes this picture from the standard modularity observed in
many real-world networks is that it automatically leads to large
assortativity: Since it is high degree nodes which form the first
cluster(s), there is a strong tendency that clusters contain nodes
with similar degrees (for previous discussions on how clustering of
nodes depends on their degree, see
e.g. \cite{ravasz_hierarchical_2003, soffer_network_2005}). Even
though the modules formed are somewhat atypical, the BRM does
demonstrate the ability of a bias for triangle formation to give rise
to community structure {\it de novo}, whereas in other models,
community structure must be put in by hand \cite{newman_social_2003}.

In the following, the clusters of tightly connected nodes created by
the BRM are called {\it clustering cores}. To visualize them, we use
what we call $q${\it-clique adjacency plots} (qCAPs) in the
following. A $q$-clique adjacency plot is based on an integer-valued
$N\times N$ matrix $T^q_{ij}$ called the $q$-clique adjacency
matrix. It is defined as $T^q_{ij}=0$ when there is no link between
$i$ and $j$, and otherwise as the number of $q$-cliques which this
link is part of. In other words, if $q = 3$, $T^{q=3}_{ij}$ is
non-zero only when $i$ and $j$ are connected, and in the $3$CAP case
it counts the number of common neighbors.  $T^q_{ij}$ can be
considered a proximity measure for nodes: linked nodes with many
common neighbors are likely to belong to the same community. Similar
proximity measures between nodes which depend on the similarity of
their neighborhoods have been used in
\cite{Ravasz_2002,leicht_2006,Ahn_2009}. To visualize $T_{ij}$, we
first rank the nodes and then plot for each pair of ranks a pixel with
corresponding color or gray scale. Possible ranking schemes are by
degree, by the number of triangles attached to the node, or by
achieving the most simple looking, block diagonalized, $q$-clique
adjacency.

Examples for the Yeast degree sequence are given in
Fig.~\ref{Figure8}. The four rows, descending from the top, show the
3CAP for typical members of the BRM ensemble before the first jump and
after the first, second, and third jumps. The plots in the left column
show the ranking done by the degrees of the nodes. The plots on the
right show the same matrices after ``diagonalization", with the nodes
forming the first cluster placed in the top ranks, followed by the
nodes forming the second cluster, and the nodes forming the third
cluster. Only the relevant parts of the 3CAPs are shown: nodes with
lower ranks do not play any substantial role except for very large
values of $\beta$. We notice several features:

\begin{itemize}
\item Not all highest degree nodes participate in the first clustering
  cores. Obviously, the selection of participating nodes is to some
  degree random, and when sufficiently many links are established they
  are frozen and cannot be changed easily later. This agrees with our
  previous observation that the positions of the jumps change
  unsystematically with details like the random number sequence or the
  speed with which $\beta$ is increased.
\item Clustering cores that have been formed once are not modified
  when $\beta$ is further increased.  Again this indicates that
  existing cores are essentially frozen.
\item Clustering cores corresponding to different steps do not
  overlap.
\end{itemize}
All three points are in perfect agreement with our previous finding
that hysteresis effects are strong and that structures which have been
formed once are preserved when $\beta$ is increased further.

From other examples (and from later jumps for the same Yeast sequence)
we know that the last two items in the list are not strictly correct
in general, although changes of cores and overlap with previous cores
do not occur often. Thus the results in Fig.~\ref{Figure8} are too
extreme to be typical. When a clustering core is formed, most of the
links connected to these nodes will be saturated, and the few links
left over will not have a big effect on the further evolution of the
core. 


We find that as $\beta^-$ decreases, the clustering cores persist well
below the value of $\beta^+$ at which they were created (not shown
here). This shows again that once a link participates in a large
number of triangles, it is very stable and unlikely to be removed
again.

$3$-clique adjacency plots are also useful for analyzing empirical
networks, independent of any rewiring null model, to help visualize
community structure. While nodes in different communities often are
linked, these links between communities usually take part in fewer
triangles than links within communities. Thus simply replacing the
standard adjacency matrix by the $3$clique adjacency matrix should help discover and
highlight community structure \cite{Ravasz_2002,leicht_2006,Ahn_2009}.

\begin{figure}
\includegraphics[scale=.25]{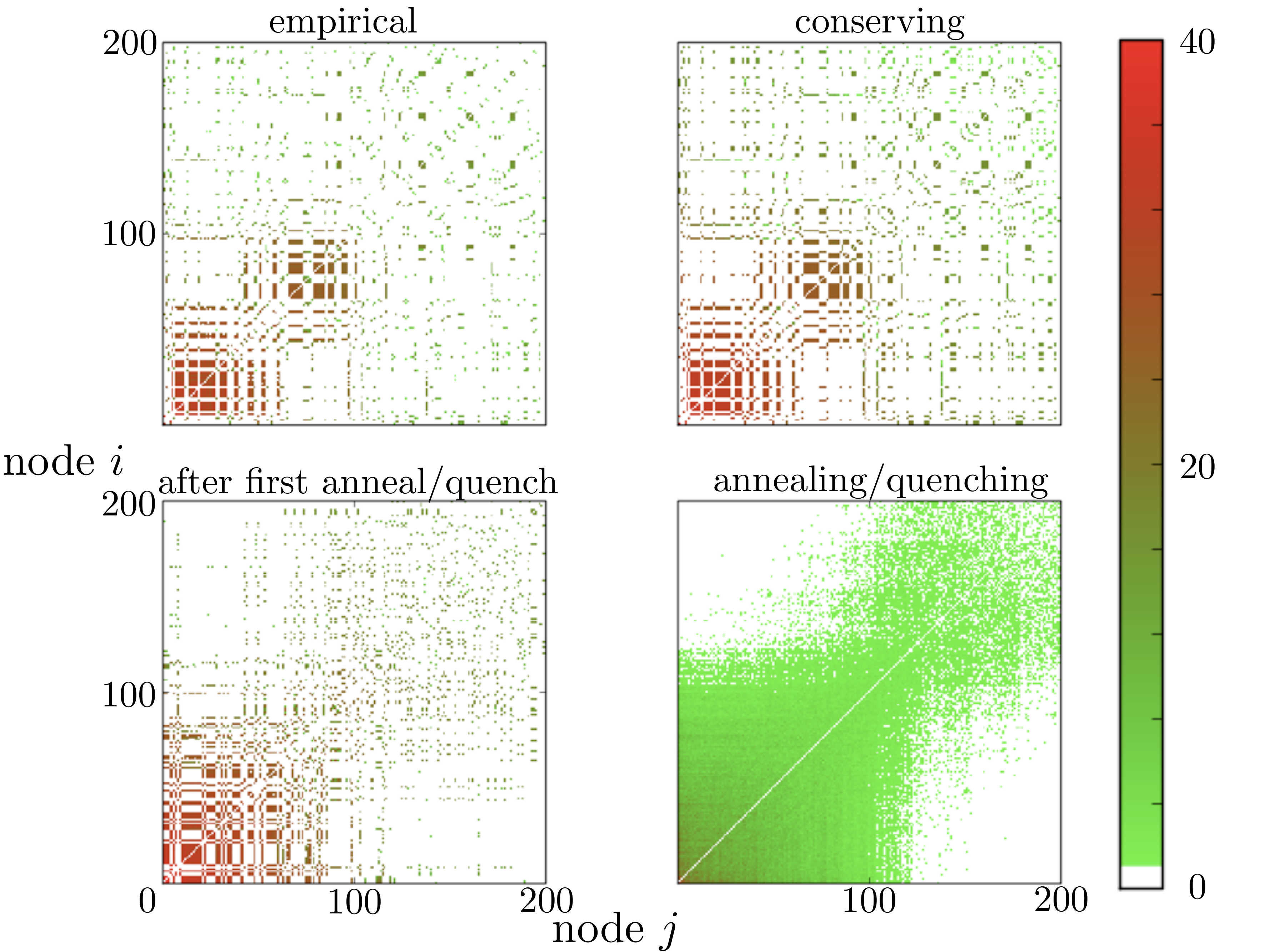}
\caption{\label{Figure9} (color online) Parts of 3CAPs for the real
  yeast protein-protein interaction network of
  \cite{gavin_functional_2002}, for a typical network of the
  ``triangle conserving" ensemble with no annealing, for a network
  obtained after an ``annealing" period with $\beta=0$ and a
  subsequent quench with $\beta\neq 0$ using `triangle conserving"
  rewirings \cite{milo_network_2002}, and for an ensemble obtained by
  500 of such annealing/quenching alternations.}
\end{figure}

\begin{figure}
\includegraphics[scale=.25]{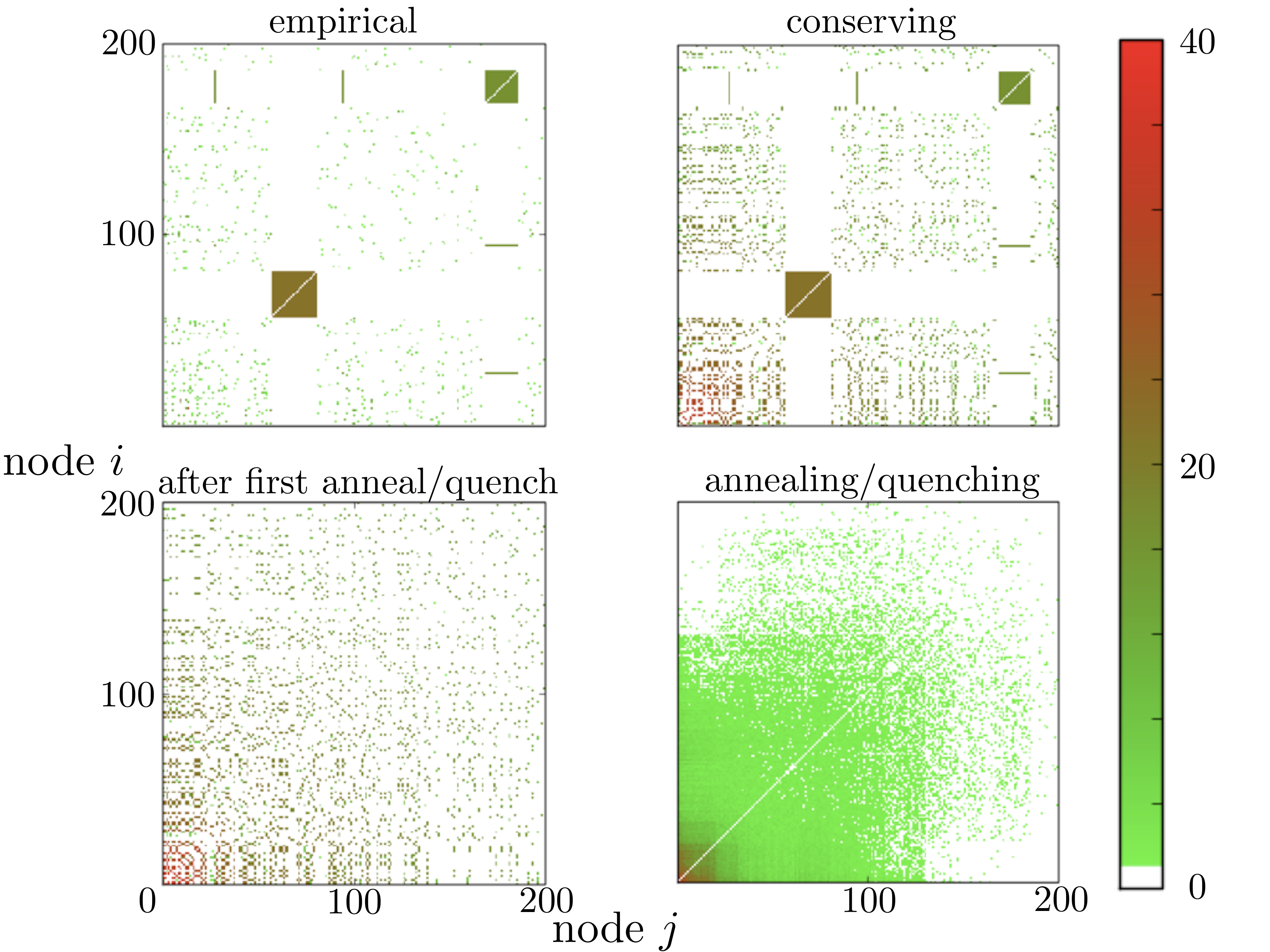}
\caption{\label{Figure10} (color online) Analogous to
  Fig.~\ref{Figure9}, but for the real high energy physics
  collaboration network and for the HEP degree sequence,
  respectively.}
\end{figure}

In the top left panel of Figs \ref{Figure9} and \ref{Figure10} we show
parts of the 3CAPs for the yeast protein-protein
interaction and HEP networks respectively. In both cases,
nodes are ranked by degree. We see that the triangles are mostly
formed between strong hubs, as we should have expected. But clustering
in the real networks does not strictly follow the degree pattern, in
the sense that some of the strongest hubs are not members of prominent
clusters. This shows again that real networks often have features
which are not encoded in their degree sequence, and that a null model
entirely based on the latter will probably fail to reproduce these
features. We see also that links typically participate in {\it many}
triangles, if they participate in at least one. This is in contrast to
a recently proposed clustering model, which assumes that each link can
only participate in a single triangle \cite{newman_random_2009}.

\subsection{Triangle conserving null models}
\label{sec:null}

In the previous subsection we considered the case where the bias is
``unidirectional". In contrast to this, Milo {\it et al.}
\cite{milo_network_2002} considered the case where the bias tends
to increase the number of triangles when it is below a number
$n_{\Delta,0}$, but pushes it {\it down} when it is above. In this way
one neither encounters any of the jumps discussed above nor any
hysteresis. But that does not mean that the method is not plagued by
the same basic problem, i.e. extreme sluggish dynamics and effectively
broken ergodicity.

In the most straightforward
implementation of triangle conserving rewiring with the Hamiltonian 
$H_{\rm Milo}$ \cite{milo_network_2002} one first estimates during
preliminary runs a value of $\beta$ which is sufficiently large so
that $n_\Delta$ fluctuates around $n_{\Delta,0}$. Then one starts with
the original true network and rewires it using this $\beta$, {\it
without first `annealing' it} to $\beta=0$.  The effect of this is
seen in the top right panels of Figs. \ref{Figure9} and \ref{Figure10}. 
In both cases, the $3$CAPs shown were obtained after $>10^9$ attempted
swaps. At $\beta=0$, this number would have been much more
than enough to equilibrize the ensemble. But for the large values of
$\beta$ needed for these plots ($\beta=1.5$ for Yeast, and $\beta=
2.4$ for HEP), few changes from the initial configurations are
seen. This is particularly true for the strongest clusters existing in
the real networks. Triangles not taking part in these clusters change
more rapidly, but are also less important.

Thus we see a pitfall inherent in triangle conserving rewiring: when
the bias is strong enough to push the number of triangles in the
network up to the desired target number, the bias will also be large
enough that links between high degree nodes are hardly ever
randomized.

\begin{figure}
\includegraphics[scale=.25]{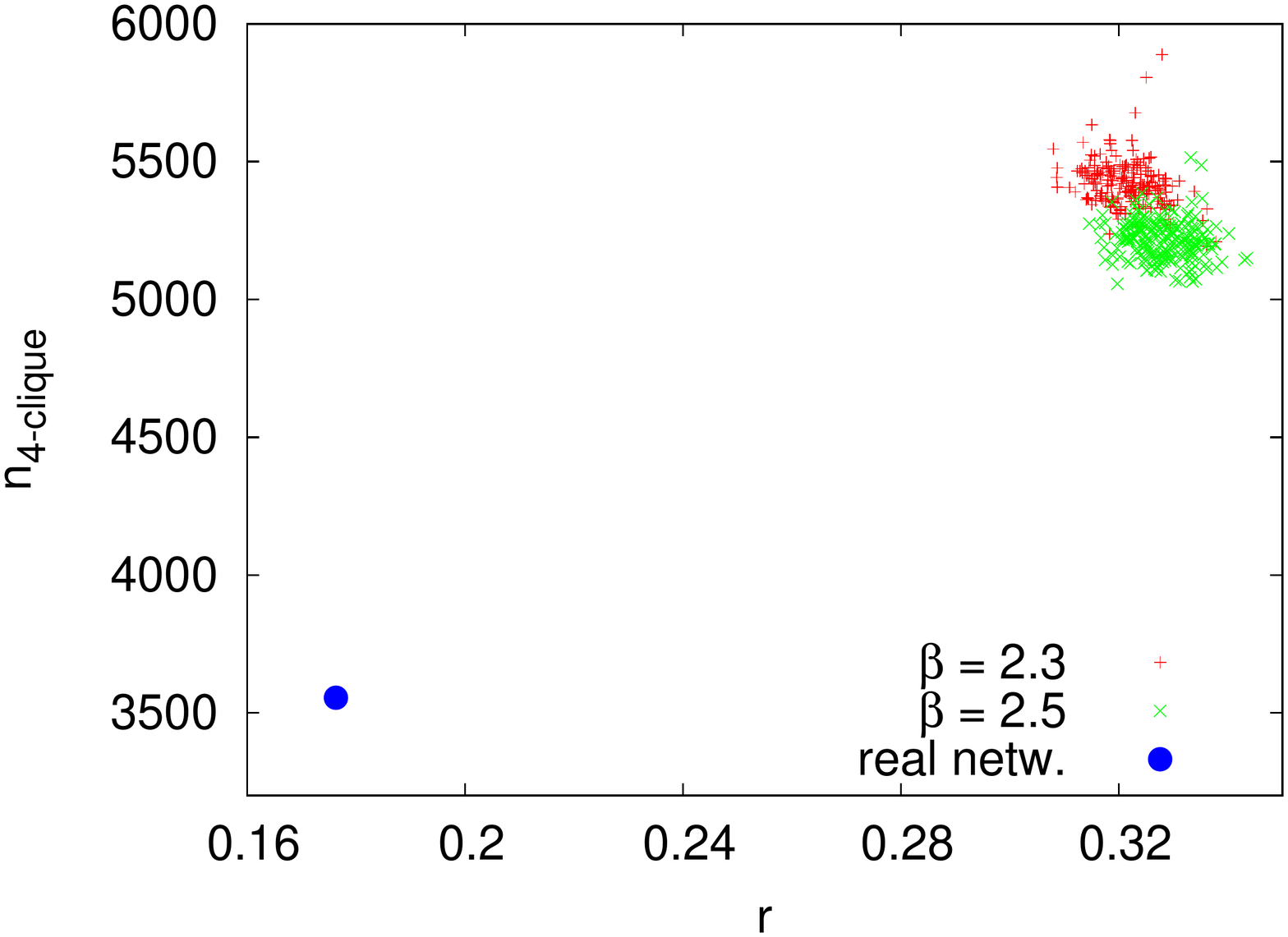}
\caption{\label{Figure11} (color online) Values of the assortativity
  $r$ and of the number of 4-cliques in the real HEP network and in
  400 members of the triangle-conserving biased ensemble.  These 400
  realizations were obtained approximately by 200 anneal/quench cycles
  with $\beta = 2.3$ and 200 cycles with $\beta = 2.5$, as described
  in the text. Notice that the results for biased simulations should
  become more exact as $\beta$ decreases towards $\beta_c \leq 2$.}
\end{figure}

As a way out of this dilemma, we can alternate epochs where we use
triangle conserving swaps with ``annealing periods" where we use
$\beta=0$. In this way we would guarantee that memory is wiped out
during each annealing period (see the lower left panels in
Figs. \ref{Figure9} and \ref{Figure10}), and each ``quenching epoch"
would thus contribute essentially one independent configuration to the
ensemble. After many such cycles we would obtain an ensemble which
looks much more evenly sampled (lower right panels in
Figs. \ref{Figure9} and \ref{Figure10}), although even then we can not
be sure that it really represents the equilibrium ensemble for the
Hamiltonian $H_{\rm Milo}$. Apart from the last caveat, the method
would presumably be too slow for practical applications where high
accuracy and precise variances of ensemble observables are needed, 
since one needs one entire cycle per data point.
But it can be useful in cases where it is
sufficient to estimate fluctuations roughly, and where high precision
is not an issue. To illustrate this, we present in Fig.~\ref{Figure11}
results for the HEP network where we made 200 anneal/quench cycles for
two different values of $\beta$ ($\beta = 2.3$ and $\beta = 2.5$). In
each cycle the quenching was stopped when the number of triangles
reached the value of the real network, and the values of $r$ and of
the number of 4-cliques was recorded. We see from Fig.~\ref{Figure11}
that these values scatter considerably, but are in all cases far from
the values for the real network. Thus the ensemble is a poor model for
the real HEP network. We also see from Fig.~\ref{Figure11} that $r$
and $n_{\rm 4-clique}$ depend slightly on $\beta$ (as was expected),
but not so much as to invalidate the above conclusion.

\section{Conclusion}
\label{sec:conc}
In highly clustered networks -- and that means for most real world
networks -- most of the clustering is concentrated amongst the
highest degree nodes. The Strauss model correctly pointed to an
important feature: clustering tends to be cooperative. Once many
triangles are formed in a certain part of the network, they help in
forming even more. Thus, clustering cannot be smoothly and evenly
introduced into a network; it is often driven by densely
interconnected, high-degree regions of the network. In triangle biased
methods these high-degree regions can emerge quite suddenly and
thereafter prove quite resistant to subsequent randomization.

The biased rewiring model studied in the present paper is of
exponential type, similar to the Strauss model, with the density of
triangles controlled by a `fugacity' or inverse `temperature'
$\beta$. However, we prevent the catastrophic increase of connectivity
at the phase transition of the Strauss model by imposing a fixed
degree sequence.  Yet there is still a first order transition for
homogeneous networks, i.e. those with fixed degree. In the phase with 
strong clustering (large fugacity / low temperature), the configuration 
is basically a collection of disjoint $k-$cliques.

If the degree sequence is not trivial, the formation of {\it
  clustering cores} can no longer happen at the same $\beta$ for
different parts of the network. Thus the single phase transition is
replaced by a sequence of discrete and discontinuous jumps, which
resemble both first order transitions and Barkhausen jumps. As in the 
real Barkhausen phenomenon, frozen randomness is crucial for the 
multiplicity of jumps. There, each 
jump corresponds to a {\it flip} of a spin cluster {\it already defined} 
by the randomness -- at least at zero temperature 
\cite{Perkovic_1995,Zapperi_1997}. In the present case, however, 
each jump corresponds 
to the {\it creation} of a cluster whose detailed properties are not 
fixed by the quenched randomness (the degree
sequence), but depend also on the `thermal' (non-quenched) noise.

As in any first order phase transition, our model shows strong
hysteresis. Clustering cores, once formed, are extremely stable and
cannot be broken up easily later.  This limits its usefulness 
as a null model, even if it is treated numerically such that the 
phase transition jumps do not appear
explicitly, as in the version of \cite{milo_network_2002}. Because of
the very slow time scales involved, Monte Carlo methods cannot
sample evenly from these ensembles. Care should be
taken to demonstrate that results found using them are
broadly consistent across various sampling procedures.

The spontaneous emergence of clustering cores in the BRM does suggest
that triangle bias can give rise to community structure in networks,
without the need to define communities {\it a priori}, thanks to the
cooperativity of triangle formation.

Together with jumps in the number of triangles (i.e. in the clustering
coefficient), there are also jumps in all other network properties at
the same control parameter positions.  In particular, we found jumps
in the number of $k-$cliques with $k>3$, in the modularity, and in the
assortativity. This immediately raises the question whether the model
can be generalized so that a different fugacity is associated to each
of these quantities. For assortativity, this was proposed some time
ago by Newman \cite{newman_assortative_2002}.  With the present
notation, biased rewiring models with and without target triangle
number $n_{\Delta,0}$ and target assortativity $r_0$ are given by the
Hamiltonians
\be
   H_{\rm Milo}(G;\beta,\gamma) = \beta |n_\Delta(G) - n_{\Delta,0} |
   + \gamma |r(G) - r_0 |
     \label{CandRHam1}
\ee
and
\be
   H_{\rm BRM}(G;\beta,\gamma) = - \beta n_\Delta(G) -\gamma r(G),
     \label{CandRHam2}
\ee
respectively, where $\gamma$ is the fugacity associated to the
assortativity.  It is an interesting open question whether such a
model might lead to less extreme clustering and thus might be more
realistic. First simulations \footnote{D. V. Foster {\it et al.}, in
  preparation} indicate that driving assortativity leads to smooth
increases of all other quantities without jumps. The reason for that
seems to be that the basic mechanism leading to increased
assortativity -- the replacement of existing links by links between
similar nodes -- is not cooperative, but further studies are needed.

As Newman remarked in \cite{newman_random_2009}, clustering in
networks ``has proved difficult to model mathematically." In that
paper he introduced a model where each link can participate in one
triangle at most. In this way, the phase transitions seen in the
Strauss model and in the present model are avoided. However, in the
real-world networks studied here we found that the number of triangles
in which a link participates is broadly distributed, suggesting that
the Newman model \cite{newman_random_2009} may not be realistic for
networks with significant clustering. Indeed, specifying for each link
the number of triangles in which it participates adds valuable
information to the adjacency matrix (which just specifies whether the
link exists or not). The resulting `$3$ clique adjacency plots'
revealed structures which would not have been easy to visualize
otherwise and are useful also in other contexts.  Thus, in contrast to
what is claimed in \cite{newman_random_2009}, the quest for realistic
models for network clustering is not yet finished.

\bibliography{references.bib}

\end{document}